\documentclass{article}
\usepackage{spconf,amsmath,amssymb,amsfonts,graphicx}
\usepackage{cite}
\usepackage{textcomp}
\usepackage{xcolor}
\usepackage{multirow}
\usepackage{booktabs}
\usepackage{comment}
\usepackage{hyperref}

\makeatletter
\renewcommand{\footnotesize}{\@setfontsize\footnotesize\@ixpt{11}}
\makeatother
\def\BibTeX{{\rm B\kern-.05em{\sc i\kern-.025em b}\kern-.08em
    T\kern-.1667em\lower.7ex\hbox{E}\kern-.125emX}}

\newcommand{\dataset}{TVPara}

\title{TVPara: Paralinguistic Captioning of Expressive TV-Series Speech in Complex Acoustic Environments}

\name{Ziyang Zhang$^{1,2}$, Wenyi Yu$^{1}$, Wen Wu$^{2}$, Xuenan Xu$^{2}$, Chao Zhang$^{1,2}$}
\address{$^{1}$Tsinghua University, China \quad $^{2}$Shanghai Artificial Intelligence Laboratory, China}

\begin{document}
\ninept
\maketitle

\begin{abstract}
Large audio-language models (LALMs) struggle to parse paralinguistics (\textit{e.g.}, emotion, pace, pitch) in speech that combines intense emotion with complex acoustic environments. Strengthening this ability calls for training data that reflects such conditions. However, existing corpora rely heavily on studio-recorded speech, offering limited emotional and acoustic diversity. To address this gap, we introduce \textbf{\dataset{}}\footnote{Audio samples are available at \url{https://windowso.github.io/TVPara/}. We will release the dataset annotations and related code.}, a paralinguistic captioning dataset built from TV series. Speech in TV series often carries diverse emotional expression and co-occurs with scene-specific background sound, providing expressive emotion and complex acoustic environments at scale. Constructed via LALM pre-annotation and human correction, \dataset{} includes two tasks: Single-Utterance ($\sim$492h) and Multi-Utterance ($\sim$213h) Paralinguistic Captioning. Fine-tuning open-source LALMs on \dataset{} yields an average relative improvement of 31.26\% in paralinguistic understanding under complex acoustic environments. The gains further generalize to an unseen TV series and to cleaner, cross-domain speech.
\end{abstract}

\begin{keywords}
paralinguistic captioning, large audio-language models, speech dataset
\end{keywords}

\section{Introduction}
\label{sec:intro}

\begin{figure}[t]
  \centering
  \includegraphics[width=\linewidth]{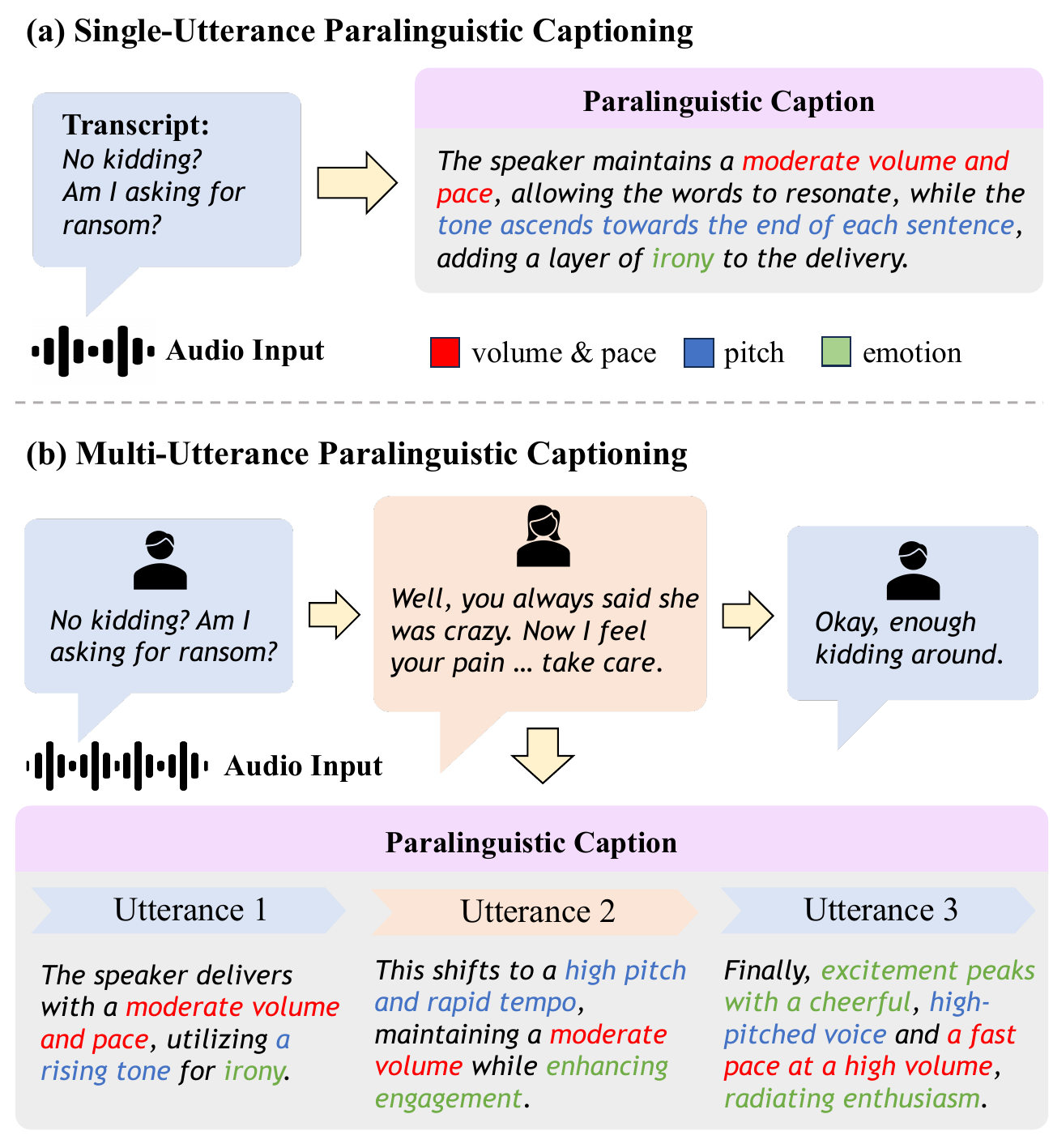}
      \caption{Overview of the two \dataset{} tasks. (a) Single-Utterance Paralinguistic Captioning describes the paralinguistic attributes of a single utterance, including emotion, pace, pitch, and volume. (b) Multi-Utterance Paralinguistic Captioning describes a sequence of adjacent utterances in order, with emphasis on how these attributes change over time (\textit{e.g.}, from irony to excitement).}
  \label{fig:task_overview}
\vspace{-0.5cm}
\end{figure}

Human speech contains rich paralinguistic information, such as emotion, pace, pitch, and volume.
Recently, large audio-language models (LALMs)~\cite{chu2023qwen, Qwen2.5-Omni, ding2025kimi, huang2025step} have shown strong generalization across speech tasks~\cite{yang2021superb, yang2024airbench}, including the ability to describe such paralinguistic information using natural language text~\cite{tangsalmonn, hurst2024gpt, chu2024qwen2}. Their captions nevertheless become coarse and inaccurate on audio that combines intense, expressive emotion with complex acoustic environments, such as background noise and sound effects~\cite{chenlisten,leemdescribe}. Both properties are common in expressive real-world audio~\cite{lotfianmsppodcast}. Closing this gap calls for training data in which the two properties naturally co-occur.

However, most existing corpora are acoustically clean: they are recorded in studios~\cite{nguyen23_interspeech, wang2020mead, richter24_interspeech, zhou2021esd}, re-annotated from clean speech libraries~\cite{zen19_interspeech, Emilia, lu24c_interspeech, TextrolSpeech, SpeechCraft, kawamura24_interspeech, diwan-etal-2025-scaling, borisov25_ssw}, or synthesized with text-to-speech~\cite{guo2023prompttts}, and thus rarely contain both intense emotion and background sound. The few collected in the wild, from podcasts~\cite{busso2025msp} to TV series~\cite{poria2019meld, zhao2022m3ed}, do contain background sound but provide only categorical or dimensional labels rather than free-form captions. As a result, corpora that combine expressive emotion and complex acoustic environments with fine-grained paralinguistic annotation are still lacking.

To bridge this gap, we construct \textbf{\dataset{}}, a paralinguistic captioning dataset built from TV series and annotated with fine-grained free-form captions. Speech in TV series often carries diverse emotional expression and co-occurs with scene-specific background sound, providing expressive emotion and complex acoustic environments at scale. 
We define two tasks (Fig.~\ref{fig:task_overview}): 1) Single-Utterance Paralinguistic Captioning ($\sim$492h), which describes the paralinguistic attributes of a single utterance; 2) Multi-Utterance Paralinguistic Captioning ($\sim$213h), which describes a sequence of adjacent utterances in order, with emphasis on how these attributes change over time.
Most existing corpora annotate each utterance in isolation and leave this task uncovered.
For annotation, we design a human-in-the-loop pipeline that combines automatic pre-annotation by LALMs with fine-grained human correction to ensure label accuracy.
Fine-tuning open-source LALMs on \dataset{} yields an average relative improvement of 31.26\% in paralinguistic understanding under complex acoustic environments. The gains further generalize to an unseen TV series and to cleaner, cross-domain speech.

The main contributions of this work are summarized as follows:
\begin{itemize}
    \item We introduce \textbf{\dataset{}}, a paralinguistic captioning dataset that pairs expressive emotion and complex acoustic environments with fine-grained free-form captions.
    \item We propose Multi-Utterance Paralinguistic Captioning, a task that captures how paralinguistic attributes change across adjacent utterances.
    \item Fine-tuning open-source LALMs on \dataset{} improves paralinguistic understanding under complex acoustic environments, with gains that generalize to an unseen TV series and to cleaner cross-domain speech.
\end{itemize}

\section{The \dataset{} Dataset}
\label{sec:dataset}

\dataset{} is built from 19 English-language TV series covering sitcoms (\textit{e.g.}, \textit{Friends}), dramas (\textit{e.g.}, \textit{Billions}), crime series (\textit{e.g.}, \textit{Poirot}), animation (\textit{The Simpsons}), and documentaries (\textit{e.g.}, \textit{Through the Wormhole})\footnote{Refer to \url{https://windowso.github.io/TVPara/} for the detailed list of TV series. The audio is clipped from the shows' official YouTube channels. We will release only the paralinguistic annotations together with the source video links and clip timestamps.}. It contains two tasks (Fig.~\ref{fig:task_overview}). Single-Utterance Paralinguistic Captioning describes the emotion, pace, pitch, and volume of one utterance. Multi-Utterance Paralinguistic Captioning takes several adjacent utterances from the same episode, spoken by one or more speakers, and describes them in order in a single caption, with emphasis on how these attributes change across utterances. Table~\ref{tab:statistics} summarizes the scale of both tasks.

\subsection{Emotion and Background Sound}
\label{subsec:characteristics}

\begin{figure}[t]
  \centering
  \includegraphics[width=\linewidth]{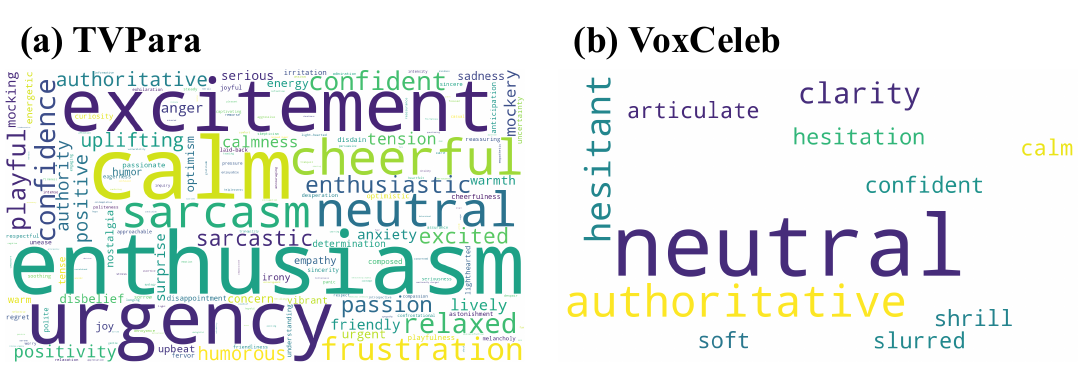}
  \caption{Emotion word clouds of (a) \dataset{} and (b) VoxCeleb. VoxCeleb is dominated by neutral and calm states, whereas \dataset{} covers many expressive emotions (\textit{e.g.}, frustration, sarcasm, anger).}
  \label{fig:wordcloud}
\vspace{-0.2cm}
\end{figure}

Speech in these TV series carries more varied emotion than cleaner speech such as VoxCeleb~\cite{nagraniy2017voxceleb} (Fig.~\ref{fig:wordcloud}), and it often co-occurs with background sound. Table~\ref{tab:characteristics} quantifies both properties: 67.2\% of \dataset{} clips carry a non-neutral emotion, compared with 2.9\% of VoxCeleb clips, and \dataset{} clips contain background sound\footnote{Background sound is detected with BEATs~\cite{chen2023beats}.} about 4.5 times as often as VoxCeleb clips. Beyond single clips, emotion also changes frequently across utterances: in multi-utterance clips, 55.5\% of adjacent utterances differ in emotion category. A single utterance cannot capture such changes, and a model that only sees single utterances does not learn to describe them, which motivates the Multi-Utterance Paralinguistic Captioning task.

\subsection{Annotation Pipeline}
\label{subsec:pipeline}

\begin{figure*}[t]
  \centering
  \includegraphics[width=\linewidth]{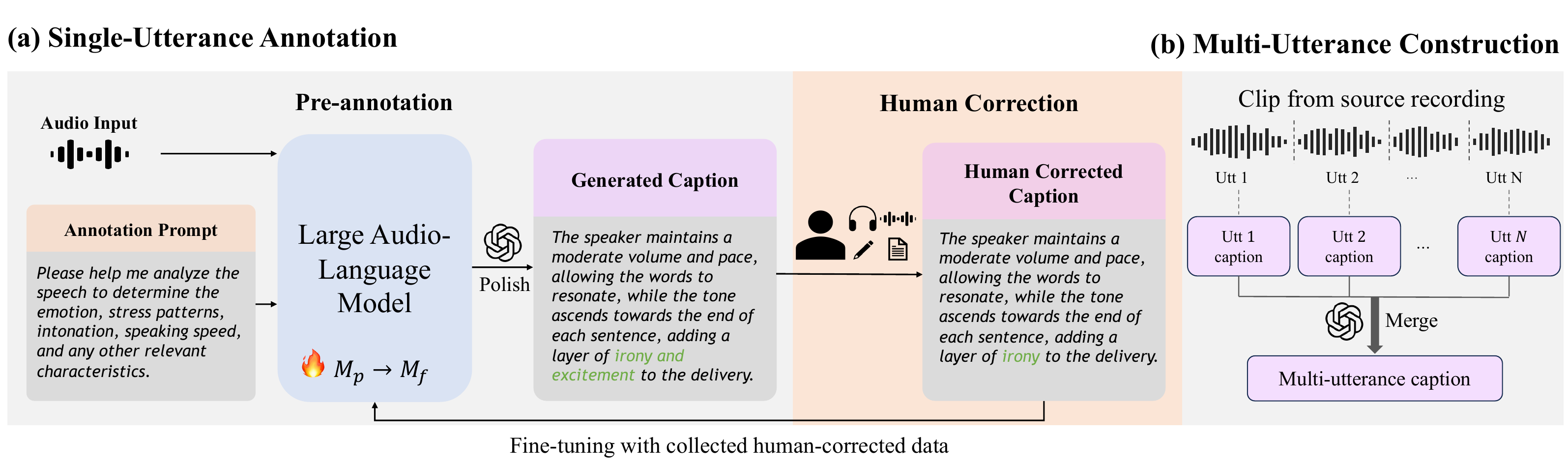}
  \caption{Annotation pipeline of \dataset{}. (a) \textbf{Single-Utterance Annotation}: each clip is pre-annotated by an LALM from the audio and a prompt, the caption is polished by an LLM, and it is then corrected by human annotators, who remove hallucinated content (\textit{e.g.}, emotions that the audio does not support) and add omitted content. The corrected data are used to fine-tune the pre-annotation model from $M_p$ to $M_f$, which then pre-annotates the remaining audio. (b) \textbf{Multi-Utterance Construction}: a clip covering $N$ adjacent utterances is cut from the source recording, and the corrected captions of the single-utterance clips within it are merged by an LLM into one caption in time order.}
  \label{fig:annotation_pipeline}
\vspace{-0.2cm}
\end{figure*}

\textbf{Single-Utterance Annotation:} Each episode's audio (about one hour long) is split into clips of a few seconds with voice activity detection (VAD)~\cite{Silero_VAD}. The clips are then annotated with the human-in-the-loop pipeline in Fig.~\ref{fig:annotation_pipeline}(a). An LALM first generates a pre-annotation for each clip\footnote{We use SALMONN-13B for pre-annotations. The checkpoint can be found at \url{https://huggingface.co/tsinghua-ee/SALMONN}.}, and a large language model (LLM) then polishes the wording of each pre-annotation before it is passed to human annotators\footnote{GPT-4o-mini is used for this step.}. Professional annotators from a commercial audio-annotation company then correct each caption against the audible cues: they remove hallucinated content such as emotion labels that the audio does not support, add attributes that the pre-annotation omits, and revise descriptions that do not match the audio. An initial subset of about one hour is corrected independently by two annotators, who agree on whether a pre-annotation needs correction for 87\% of its clips; each of the remaining clips is then corrected by one annotator. After $\sim$50 hours of corrected data, the pre-annotation LALM is fine-tuned on that data and then used to pre-annotate the remaining audio. Annotators initially changed about 38\% of the pre-annotations; after the pre-annotation model was fine-tuned on corrected data, this rate fell to about 17\%.

\begin{table}[t]
  \caption{Statistics of the Single- and Multi-Utterance Paralinguistic Captioning tasks of \dataset{}.}
  \label{tab:statistics}
  \centering
  \begin{tabular}{lcc}
    \toprule
    \textbf{Statistic} & \textbf{Single} & \textbf{Multi} \\
    \midrule
    Total audio clips & 256,212 & 45,402 \\
    Total duration (h) & 492 & 213 \\
    Avg. clip duration (s) & 6.91 & 16.89 \\
    Avg. caption length (words) & 32.8 & 58.1 \\
    \bottomrule
  \end{tabular}
\vspace{-0.3cm}
\end{table}

\begin{table}[t]
  \caption{Emotion and background sound statistics of \dataset{} and VoxCeleb. VoxCeleb has no multi-utterance data.}
  \label{tab:characteristics}
  \centering
  \resizebox{\linewidth}{!}{%
  \begin{tabular}{lcc}
    \toprule
    \textbf{Statistic} & \textbf{\dataset{}} & \textbf{VoxCeleb} \\
    \midrule
    Clips with non-neutral emotion & 67.2\% & 2.9\% \\
    Clips with background sound & 36.2\% & 8.0\% \\
    Adjacent utterances with emotion change & 55.5\% & -- \\
    \bottomrule
  \end{tabular}}
\vspace{-0.3cm}
\end{table}

\begin{figure*}[t]
  \centering
  \includegraphics[width=0.97\linewidth]{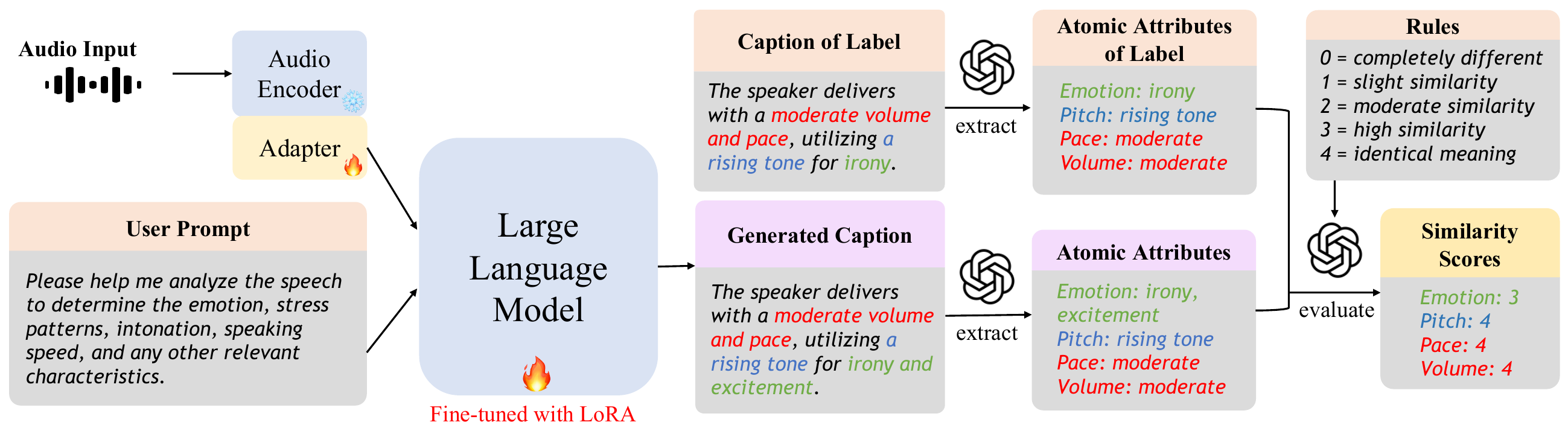}
  \caption{Illustration of the inference and LLM-as-a-Judge evaluation pipeline. During inference, the audio and user prompt are processed by the LALM to generate a paralinguistic caption. For evaluation, an LLM extracts atomic attributes (emotion, pace, pitch, volume) from both the generated caption and the ground-truth label. An LLM evaluator then compares these aligned attributes, assigning similarity scores on a discrete scale from 0 to 4.}
  \label{fig:infer_eval}
\vspace{-0.2cm}
\end{figure*}

\textbf{Multi-Utterance Construction:} Multi-utterance data are derived from the single-utterance annotations, as shown in Fig.~\ref{fig:annotation_pipeline}(b). We cut longer clips directly from the source recordings, each covering $N$ ($N \in [3, 5]$) adjacent utterances from the same episode, so the acoustic continuity of the original scene is preserved. Using the timestamps, we locate the single-utterance clips within each multi-utterance clip and retrieve their human-corrected captions, which an LLM then merges into one caption in time order, with emphasis on how the attributes change across utterances\footnote{GPT-4o-mini is used for this step.}. A 5-hour sample of the merged captions was given to the annotators for correction, and they changed only about 4\% of them, so the remaining multi-utterance captions are used without manual correction.

\subsection{Data Splits}
\label{subsec:splits}

\dataset{} provides two test sets. For the unseen-series test set, one whole TV series is held out from training, so none of its speakers appear in the training set. One episode (about 1 hour) of this series is used as the test set. The other TV series are split into the training set and the hold-out test set: one continuous portion of each episode is randomly selected for testing, and the rest of the episode is used for training. Both the single- and multi-utterance test clips are cut from this portion, so the training and test sets share no audio. The hold-out test set contains about 5 hours of audio, leaving about 487 hours for training.

\section{Experimental Setup}
\label{sec:experiments}

\subsection{Training}
\label{subsec:training}
Two open-source LALMs are fine-tuned: SALMONN~\cite{tangsalmonn} and Qwen2.5-Omni\footnote{\url{https://huggingface.co/Qwen/Qwen2.5-Omni-7B}}~\cite{Qwen2.5-Omni}. For both models, the audio encoder is frozen, the cross-modal adapter is fine-tuned, and the LLM is fine-tuned with Low-Rank Adaptation (LoRA)~\cite{hu2022lora} ($r=8$, $\alpha=32$). Both models are trained with Supervised Fine-Tuning (SFT) on NVIDIA H200 GPUs, using shuffled single- and multi-utterance training data so that one checkpoint serves both tasks. Training runs for 10 epochs with a batch size of 24 and a cosine learning-rate schedule with a peak learning rate of $3 \times 10^{-6}$ and 5\% warmup. SALMONN is trained with its official codebase\footnote{\url{https://github.com/bytedance/SALMONN}}, and Qwen2.5-Omni with \texttt{ms-swift}\footnote{\url{https://github.com/modelscope/ms-swift}}.

\begin{table*}[h]
  \caption{Evaluation results of paralinguistic captioning on the four attributes (emotion, pace, pitch, and volume; scored from 0 to 4) on the \dataset{} hold-out test set. The better result for each base model is in \textbf{bold}.}
  \label{tab:main_results}
  \centering
  \vspace{2pt}
  \begin{tabular}{l l cccc c cccc}
    \toprule
    \multirow{2}{*}{\textbf{Model}} & \multirow{2}{*}{\textbf{Training}} & \multicolumn{4}{c}{\textbf{Single-Utterance Task}} && \multicolumn{4}{c}{\textbf{Multi-Utterance Task}} \\
    \cmidrule{3-6} \cmidrule{8-11}
    & & Emotion & Pace & Pitch & Volume && Emotion & Pace & Pitch & Volume \\
    \midrule
    \multirow{2}{*}{Qwen2.5-Omni}
      & Zero-shot & 0.74 & 2.02 & 1.41 & \textbf{2.43} && 0.99 & 1.29 & 1.26 & 1.12 \\
      & SFT       & \textbf{0.99} & \textbf{2.40} & \textbf{2.45} & 2.33 && \textbf{1.46} & \textbf{1.55} & \textbf{1.48} & \textbf{1.35} \\
    \midrule
    \multirow{2}{*}{SALMONN}
      & Zero-shot & 1.45 & \textbf{2.74} & 2.32 & 2.31 && 1.09 & 1.21 & 1.10 & 1.04 \\
      & SFT       & \textbf{1.52} & 2.73 & \textbf{2.44} & \textbf{3.58} && \textbf{1.60} & \textbf{1.60} & \textbf{1.70} & \textbf{1.81} \\
    \bottomrule
\end{tabular}
\vspace{-0.3cm}
\end{table*}

\begin{table*}[!t]
  \caption{Out-of-distribution evaluation on the four attributes (scored from 0 to 4). Unseen Series is the unseen-series test set (Section~\ref{subsec:splits}); VoxCeleb is the cleaner, cross-domain test set (Section~\ref{subsec:eval}, single-utterance only). The better result for each base model is in \textbf{bold}.}
  \label{tab:ood}
  \centering
  \resizebox{\textwidth}{!}{
  \begin{tabular}{l l cccc c cccc c cccc}
    \toprule
    \multirow{2}{*}{\textbf{Model}} & \multirow{2}{*}{\textbf{Training}} & \multicolumn{4}{c}{\textbf{Unseen Series (Single)}} && \multicolumn{4}{c}{\textbf{Unseen Series (Multi)}} && \multicolumn{4}{c}{\textbf{VoxCeleb (Single)}} \\
    \cmidrule{3-6} \cmidrule{8-11} \cmidrule{13-16}
    & & Emo. & Pace & Pitch & Vol. && Emo. & Pace & Pitch & Vol. && Emo. & Pace & Pitch & Vol. \\
    \midrule
    \multirow{2}{*}{Qwen2.5-Omni}
      & Zero-shot  & 0.74 & \textbf{2.21} & 1.45 & 1.47 && 1.07 & \textbf{1.36} & 1.26 & \textbf{1.11} && 0.84 & \textbf{2.24} & 1.45 & 1.80 \\
      & SFT        & \textbf{0.95} & 1.78 & \textbf{2.30} & \textbf{2.44} && \textbf{1.19} & 1.28 & \textbf{1.42} & 1.08 && \textbf{0.96} & 1.86 & \textbf{2.43} & \textbf{2.22} \\
    \midrule
    \multirow{2}{*}{SALMONN}
      & Zero-shot  & 1.35 & 2.64 & 2.08 & 1.65 && 1.17 & 1.48 & 1.30 & 1.13 && 1.25 & 2.70 & 2.02 & 1.79 \\
      & SFT        & \textbf{1.52} & \textbf{2.89} & \textbf{2.48} & \textbf{3.51} && \textbf{1.64} & \textbf{1.55} & \textbf{1.76} & \textbf{1.83} && \textbf{1.39} & \textbf{2.85} & \textbf{2.34} & \textbf{3.69} \\
    \bottomrule
  \end{tabular}
  }
\vspace{-0.3cm}
\end{table*}

\subsection{Evaluation}
\label{subsec:eval}
Models are evaluated on the \dataset{} hold-out and unseen-series test sets (Section~\ref{subsec:splits}) and on a VoxCeleb test set, which measures transfer to cleaner, cross-domain interview speech. The VoxCeleb test set contains 134 single-utterance clips, annotated with the pipeline in Section~\ref{subsec:pipeline}. Unlike \dataset{}, this set has light background and mostly neutral and calm emotions (Fig.~\ref{fig:wordcloud}).

\subsection{Metrics}
\label{subsec:metrics}
As shown in Fig.~\ref{fig:infer_eval}, an LLM-as-a-Judge approach is used for scoring. An LLM\footnote{GPT-5-mini is used for automated evaluation.} extracts the atomic paralinguistic attributes (emotion, pace, pitch, volume) from both the generated caption and the human-corrected ground-truth label, compares them attribute by attribute, and assigns each attribute a similarity score from 0 to 4, where a higher score means closer agreement with the label. The judge does not receive the audio: it only extracts and compares attributes at the text level, and the acoustic perception is supplied by the human-corrected label.

\section{Results and Analysis}
\label{sec:results}

\subsection{Main Results on \dataset{}}
\label{subsec:main_results}

Table~\ref{tab:main_results} reports the results. In the zero-shot setting, both LALMs score low on most attributes, especially on the multi-utterance task, showing that they struggle with paralinguistic captioning in complex acoustic environments. Fine-tuning on \dataset{} improves both models: the mean per-cell relative improvement of SFT over the zero-shot baseline is 31.26\%, and nearly every gain is significant under a Wilcoxon signed-rank test on per-clip scores ($p{<}0.01$).

We further verify that some gains come from audio perception rather than from a better fit to the label distribution, by computing the Spearman correlation between what the model predicts and what the audio actually contains. On the single-utterance task, SALMONN's predicted volume level (low, moderate, or high) correlates with the measured clip energy at $\rho{=}0.85$ after fine-tuning, against $0.03$ in the zero-shot setting. On the multi-utterance task, the number of utterances Qwen2.5-Omni describes correlates with the number of utterances in the clip at $\rho{=}0.83$, against $0.05$ in the zero-shot setting.

Qwen2.5-Omni fits the label distribution rather than the audio on volume: the correlation between its predicted level and the measured clip energy stays at $\rho{\approx}0.03$ before and after fine-tuning, and fine-tuning only shifts its output distribution toward the training labels. We attribute this to the capacity of the audio adapter: Qwen2.5-Omni's adapter has about 1/690 the trainable parameters of SALMONN's, so its audio tokens may change little during fine-tuning and may not encode every paralinguistic attribute. The LLM's LoRA weights then fit the targets from a representation that lacks the needed information, and for volume the model fits the label distribution instead of the audio.

\subsection{Out-of-Distribution Generalization}
\label{subsec:ood}

Table~\ref{tab:ood} shows that the gains generalize to both out-of-distribution test sets. On the unseen series, SALMONN improves on all four attributes in both tasks, and Qwen2.5-Omni improves on three of four single-utterance attributes and two of four multi-utterance attributes. The gain is larger on clips with background sound: on the unseen series, the mean per-cell relative improvement is 32.4\% on clips with background sound and 19.9\% on clips without. The learned ability thus transfers to new expressive and acoustically complex content. On the cleaner VoxCeleb speech, training on complex acoustics improves most attributes rather than degrading them: SALMONN improves on all four, and Qwen2.5-Omni on all but pace. Some gains come from audio perception rather than from a better fit to the label distribution. The label distribution shifts across sets, with most training clips labeled moderate in volume but most test clips in both out-of-distribution test sets labeled high, yet SALMONN's predicted volume level remains correlated with measured clip energy on both the unseen series (Spearman $\rho{=}0.87$) and VoxCeleb ($\rho{=}0.73$).

\section{Conclusion}
\label{sec:conclusion}

We introduced \textbf{\dataset{}}, a paralinguistic captioning dataset built from TV series, whose speech carries expressive emotion and complex acoustic environments, and defined two tasks on it: Single-Utterance and Multi-Utterance Paralinguistic Captioning. The study supports three findings. First, most existing corpora offer limited emotional and acoustic diversity and annotate each utterance in isolation. \dataset{} pairs both kinds of diversity with fine-grained captions and covers how paralinguistic attributes change across adjacent utterances. Second, the two open-source LALMs we tested show limited paralinguistic understanding on audio with intense emotion and complex background sound, and fine-tuning on \dataset{} yields an average relative improvement of 31.26\% in paralinguistic understanding under complex acoustic environments. Third, the improved ability transfers to an unseen TV series and to cleaner, cross-domain speech.

\section{Compliance with Ethical Standards}
\label{sec:ethics}
The audio in \dataset{} is clipped from publicly available videos on the official YouTube channels of the TV series. We do not distribute the audio; we release only the paralinguistic annotations together with the source video links and clip timestamps. The speech consists of scripted performances by professional actors rather than recordings collected from private individuals, so this work does not constitute human-subject research and no ethical approval was required. Annotation was done by a commercial data-annotation company, whose annotators were paid in accordance with local labor laws.

\section{Funding Acknowledgements}
\label{sec:funding}
No funding was received for conducting this study. The authors have no relevant financial or nonfinancial interests to disclose.

\bibliographystyle{IEEEbib}
\bibliography{tvpara_arxiv}

\end{document}